\documentstyle[12pt]{article}
\makeatletter
\ifcase \@ptsize
\or % mods for 11 pt
\or % mods for 12 pt
\fi
\advance\textheight by \topskip

\ifcase \@ptsize
\or % mods for 11 pt
\or % mods for 12 pt
\fi

\def\@citex[#1]#2{%
\if@filesw \immediate \write \@auxout {\string \citation {#2}}\fi
\@tempcntb\m@ne \let\@h@ld\relax \def\@citea{}%
\@cite{%
  \@for \@citeb:=#2\do {%
    \@ifundefined {b@\@citeb}%
      {\@h@ld\@citea\@tempcntb\m@ne{\bf ?}%
      \@warning {Citation `\@citeb ' on page \thepage \space undefined}}%
      {\@tempcnta\@tempcntb \advance\@tempcnta\@ne%
      \@tempcntb\number\csname b@\@citeb \endcsname \relax%
      \ifnum\@tempcnta=\@tempcntb %   Number follows previous--hold on to it
        \ifx\@h@ld\relax%
          \edef \@h@ld{\@citea\csname b@\@citeb\endcsname}%
        \else%
          \edef\@h@ld{\ifmmode{-}\else--\fi\csname b@\@citeb\endcsname}%
        \fi%
      \else%   %  non-successor--dump what's held and do this one
        \@h@ld\@citea\csname b@\@citeb \endcsname%
        \let\@h@ld\relax%
      \fi}%
    \def\@citea{,\penalty\@highpenalty\,}%
  }\@h@ld
}{#1}}
\@addtoreset{equation}{section}
\def\section{\@startsection {section}{1}{\z@}{-3.5ex plus -1ex minus
 -.2ex}{2.3ex plus .2ex}{\large\bf\centering}}
\def\subsection{\@startsection{subsection}{2}{\z@}{-3.25ex plus -1ex minus
 -.2ex}{1.5ex plus .2ex}{\sc}}
\gdef\@publabel{\hfil}
\gdef\@pubdate{\null}
\gdef\@pubnumber{\null}
\gdef\@author{\null}
\gdef\@title{\null}
\gdef\@abstract{\null}
\long\def\pubdate#1{\gdef\@pubdate{#1}}
\long\def\pubnumber#1{\gdef\@pubnumber{#1}}
\long\def\publabel#1{\gdef\@publabel{#1}}
\long\def\author#1{\gdef\@author{#1}}
\long\def\title#1{\gdef\@title{#1}}
\long\def\abstract#1{\gdef\@abstract{#1}}
\def\titlerelax{
\let\maketitle\relax
\let\settitleparameters\relax
\let\consolidatetitle\relax
\let\inittitlepage\relax
\let\finishtitlepage\relax
\let\titlepagecontents\relax
\let\multithanks\relax
\let\titlebaselines\relax
\let\@makepub\relax
\let\@maketitle\relax
\let\@makeauthor\relax
\let\@makeabstract\relax
\let\@maketitlenote\relax
\let\thanks\relax
\let\titlerelax\relax}
\def\titleclean
{\gdef\@titlenote{}
\gdef\@abstract{}
\gdef\@author{}
\gdef\@title{}
\gdef\@pubdate{}\gdef\@pubnumber{}\gdef\@publabel{}
\gdef\@dpublabel{}
}
\def\@makepub{\vbox to \z@{\hbox to \textwidth{\hfill
\@publabel \hfill
\llap{\parbox[t]{0.33\textwidth}{\raggedleft\@pubnumber}}}%
\vss}}
\def\@maketitle{\vskip 60pt \begin{center}
 {\LARGE \@title \par}
 \end{center}}
\def\@makeauthor{{%
\def\and{\smallskip {\normalsize \rm and\smallskip }}
\def\And{\medskip {\normalsize \rm and\\}\medskip}
\long\def\address##1{{\def\and{\\and\\}\medskip
				{\small \it \\##1\\}
}}
{\centering
 \vskip 3em
 \large \lineskip .75em
 \@author}
 \par}}
\def\@makedate{\vskip 1.5em
 {\raggedright \small \noindent\@pubdate \par}}
\def\@makeabstract{\vskip 1.5em
{\small
\begin{center}
{\bf ABSTRACT\vspace{-.5em}\vspace{0pt}}
\end{center}
\quotation \@abstract \endquotation}}
\def\maketitle{\titlepage
\let\footnotesize\small \setcounter{page}{0}
\@makepub
\vfil
\@maketitle
\@makeauthor
\vfil
\@makeabstract
\@thanks
\vfil
\@makedate
\if@restonecol\twocolumn \else \eject \fi
\titlerelax \titleclean
\setcounter{footnote}{0}
}
\def\square{\hbox{\vrule height2ex\kern-0.4pt
\vbox to 2ex{\hrule width2ex\vfil\hrule width2ex}\kern-0.4pt\vrule height2ex}}

{\unskip\nobreak\hfill$\Box$\end{list}}
\def\lefteqn#1{\hbox to 0pt{$\displaystyle #1$\hss}}
\def\starteqn#1#2{\hbox to 0pt{${\displaystyle #1}\textstyle #2$\hss}
\qquad\nonumber\\*&#2&}
\def\cA{{\cal A}}

\def\cF{{\cal F}}

\def\cH{{\cal H}}
\def\cI{{\cal I}}
\def\cL{{\cal L}}

\let\a=\alpha
\let\b=\beta
\let\d=\delta

\let\p=\phi

\def\bfl{{\bf l}}

\def\blank#1{}

\def\bra#1.{{\langle#1|}}
\def\cca(#1,#2){{\textstyle{#1 \atopwithdelims[] #2}}}
\def\ccb(#1,#2){{\textstyle{#1 \atopwithdelims\{\} #2}}}
\def\cdd{{\cdot}}
\def\cev#1{\langle #1 \vert}

\def\cont{\nonumber\\*&&\mbox{}}

\def\dpl{{\partial_+}}
\def\dslash{{\partial}{\kern -1.5ex\hbox{/}}}

\def\eb+#1#2{E_+^{#1}(#2)}

\def\en{\end{equation}}

\def\enn{\end{eqnarray}}

\def\eq{\begin{equation}}
\def\eqq{\begin{eqnarray}}
\def\F(#1,#2,#3,#4|#5|#6,#7,#8){\cF^{#2,#3}_{#1,#4}(#5|#6,#7,#8)}

\def\half{{1\over2}}
\def\half#1{\frac{#1}{2}}

\def\ket#1.{{|#1\rangle}}

\def\mno{{\textstyle {\circ\atop\circ}}}

\def\no{{\textstyle {\times\atop\times}}}

\def\num{\global\advance \eqnnumber by 1}
\def\num#1#2{N^{(#1)}(#2)}
\def\numa #1{(\the\eqnnumber \, #1)}

\def\path{\{\bfl\}}

\def\Tr{\mathop{\rm Tr}\nolimits}

\def\vac{\vec 0}

\def\vchi#1{\chi\!{\raise-1pt\hbox{ ${}^V_{#1}$ }} }
\def\vec#1{\vert #1 \rangle}

\def\vo(#1,#2)#3{{#1 \choose #2}_{\!#3}}

\def\WA{\mathop{\it WA}\nolimits}
\def\wan{$\WA_n$ }

\def\WD{\mathop{\it WD}\nolimits}

\def\eq{\begin{equation}}
\def\en{\end{equation}}

\def\eqq{\begin{eqnarray}}
\def\enn{\end{eqnarray}}

\newcount{\q}
\thicklines
\def\spot{\circle{2}}
\def\dit{\line(1,0){8}}

\def\dynkina#1{
\q=#1 \advance\q by -1
\put(00,0){\circle{2}}
\multiput(01,0)(10,0){\q}{\line(1,0){8}}
\multiput(10,0)(10,0){\q}{\circle{2}}
}

\def\dynkinc#1{
\q=#1 \advance\q by -2
\put(00,0){\circle{2}}
\put(10,0){\spot}
\put(01,1){\dit}
\put(01,-1){\dit}
\put(3,5){\line(1,-1){5}}
\put(3,-5){\line(1,1){5}}
\multiput(11,0)(10,0){\q}{\line(1,0){8}}
\multiput(20,0)(10,0){\q}{\circle{2}}
}

\def\dynkinb#1{
\q=#1 \advance\q by -2
\put(00,0){\circle{2}}
\put(10,0){\spot}
\put(01,1){\dit}
\put(01,-1){\dit}
\put(02,0){\line(1,-1){5}}
\put(02,0){\line(1,1){5}}
\multiput(11,0)(10,0){\q}{\line(1,0){8}}
\multiput(20,0)(10,0){\q}{\circle{2}}
}

\def\dynkind#1{
\q=#1 \advance\q by -3
\put(03,7){\spot}
\put(03,-7){\spot}
\put(10,0){\spot}
\put(04,-6){\line(1,1){5}}
\put(04,6){\line(1,-1){5}}
\multiput(11,0)(10,0){\q}{\line(1,0){8}}
\multiput(20,0)(10,0){\q}{\circle{2}}
}

\def\dynking2{
\put(0,0){\spot}
\put(10,0){\spot}
\put(2,5){\line(1,-1){5}}
\put(2,-5){\line(1,1){5}}
\put(1,0){\dit}
\put(1,1){\dit}
\put(1,-1){\dit}
}

\def\adynking2{
\put(0,0){\spot}
\put(10,0){\spot}
\put(3,0){\line(1,-1){5}}
\put(3,0){\line(1,1){5}}
\put(1,0){\dit}
\put(1,1){\dit}
\put(1,-1){\dit}
\put(11,0){\dit}
\put(20,0){\spot}
}

\def\adynkind4{
\put(0,0){\spot}
\put(10,0){\spot}
\put(2,-5){\line(1,1){5}}
\put(2,5){\line(1,-1){5}}
\put(1,0){\dit}
\put(1,1){\dit}
\put(1,-1){\dit}
\put(11,0){\dit}
\put(20,0){\spot}
}

\def\fone{			% B_3^1
\dynkind3
\put(20,0){\spot}
\put(11,1){\dit}
\put(11,-1){\dit}
\put(13,-5){\line(1,1){5}}
\put(13,5){\line(1,-1){5}}
}

\def\ftwo{			% A_5^2
\dynkind3
\put(20,0){\spot}
\put(11,1){\dit}
\put(11,-1){\dit}
\put(12,0){\line(1,1){5}}
\put(12,0){\line(1,-1){5}}
}

\def\fthr{			% C_3^1
\dynkinc3
\put(30,0){\spot}
\put(21,1){\dit}
\put(21,-1){\dit}
\put(23,0){\line(1,1){5}}
\put(23,0){\line(1,-1){5}}
}

\def\fnin{			% A_6^2
\dynkinb3
\put(30,0){\spot}
\put(21,1){\dit}
\put(21,-1){\dit}
\put(23,0){\line(1,1){5}}
\put(23,0){\line(1,-1){5}}
}

\def\ffou{			% D_5^2
\dynkinb3
\put(30,0){\spot}
\put(21,1){\dit}
\put(21,-1){\dit}
\put(22,-5){\line(1,1){5}}
\put(22,5){\line(1,-1){5}}
}

\def\ften{			% A_6^2
\dynkinc3
\put(30,0){\spot}
\put(21,1){\dit}
\put(21,-1){\dit}
\put(22,-5){\line(1,1){5}}
\put(22,5){\line(1,-1){5}}
}

\def\fsev{			% A_4^2 (i)
\dynkinc2
\put(20,0){\spot}
\put(11,1){\dit}
\put(11,-1){\dit}
\put(12,-5){\line(1,1){5}}
\put(12,5){\line(1,-1){5}}
}

\def\ffiv{			% C_2^1
\dynkinc2
\put(20,0){\spot}
\put(11,1){\dit}
\put(11,-1){\dit}
\put(12,0){\line(1,1){5}}
\put(12,0){\line(1,-1){5}}
}

\def\feig{			% A_4^2 (ii)
\dynkinb2
\put(20,0){\spot}
\put(11,1){\dit}
\put(11,-1){\dit}
\put(12,0){\line(1,1){5}}
\put(12,0){\line(1,-1){5}}
}

\def\fsix{			% D_3^2
\dynkinb2
\put(20,0){\spot}
\put(11,1){\dit}
\put(11,-1){\dit}
\put(12,-5){\line(1,1){5}}
\put(12,5){\line(1,-1){5}}
}

\def\Gone{
\begin{picture}(30,25)(-5,0)
\fone
\end{picture}
}

\def\Gtwo{
\begin{picture}(30,25)(-5,0)
\ftwo
\end{picture}
}
\def\Gthr{
\begin{picture}(40,25)(-5,0)
\fthr
\end{picture}
}
\def\Gfou{
\begin{picture}(40,25)(-5,0)
\ffou
\end{picture}
}
\def\Gfiv{
\begin{picture}(30,25)(-5,0)
\ffiv
\end{picture}
}
\def\Gsix{
\begin{picture}(30,25)(-5,0)
\fsix
\end{picture}
}
\def\Gsev{
\begin{picture}(30,25)(-5,0)
\fsev
\end{picture}
}
\def\Geig{
\begin{picture}(30,25)(-5,0)
\feig
\end{picture}
}
\def\Gnin{
\begin{picture}(30,25)(-5,0)
\fnin
\end{picture}
}
\def\Gten{
\begin{picture}(30,25)(-5,0)
\ften
\end{picture}
}

\def\Table{\@ifnextchar[{\@Table}{\@Table[h]}}
\def\@Table[#1]{\par\def\@captype{table}%
\vbox\bgroup\hsize\columnwidth\@parboxrestore}
\def\endTable{\par\vskip\z@\egroup}
\def\vextend{\hskip -.5\arrayrulewidth \vrule \@width
\arrayrulewidth \@height 5pt \hskip -.5\arrayrulewidth}
\def\emptycol#1{\omit\vextend\hfill\vextend\global\@tempcnta#1\relax
\ifnum\@tempcnta>1\let\next=\@emptycol\else\let\next=\ignorespaces\fi\next}
\def\@emptycol{&\omit\hfill\vextend\global\advance\@tempcnta by -1
\ifnum\@tempcnta>1\let\next=\@emptycol\else\let\next=\ignorespaces\fi\next}
\def\sline#1{\emptycol{#1}\\\hline\emptycol{#1}\\}

\makeatother
\begin{document}

\pubnumber{{Imperial/TH/91-92/17} \\ DAMTP 92-07 \\ DUR-CPT 92-01}
\pubdate{Feb 1992}
\title{Duality in Quantum Toda theory and W-algebras}
\author{
H. G. KAUSCH\thanks{Email: {\tt H\_G\_KAUSCH@V1.PH.IC.AC.UK}}
\address{
Blackett Laboratory, Imperial College,\thanks{Present address}\\
London, SW7~2BZ, U.K. \\
\And
Department of Applied Mathematics and Theoretical Physics,\\
University of Cambridge, Silver Street,     \\
Cambridge, CB3 9EW, U.K.
}
\bigskip
G. M. T. WATTS\thanks{Email: {\tt G.M.T.Watts@UK.AC.DURHAM}}\address{
Department of Mathematical Sciences, \\
University of Durham, South Road,
Durham, DH1 3LE, U.K.}
}

\abstract{
We consider Quantum Toda theory associated to a general Lie algebra.
We prove that the conserved quantities in both conformal and affine
Toda theories exhibit duality interchanging the Dynkin diagram and its
dual, and inverting the coupling constant. As an example we discuss
the conformal Toda theories based on $B_2,B_3$ and $G_2$ and the
related affine theories.
}

\maketitle

\section{Introduction}

\def\wdn{$\WD_n$ }

It was noticed some time ago that there is a relation between the
conformal Toda theory based on a simply-laced algebra $g$ with
coupling constant $\beta$, and $1/\beta$. This has also been extended
to the case of a non-simply-laced Lie algebra $g$; here the conformal
Toda theory given by $g$, with coupling constant $\beta$ is related to
the Toda theory of $g^\vee$ with coupling constant $1/\beta$.
(The Dynkin diagram of $g^\vee$ is obtained by changing the directions
of all the arrows which appear in the
Dynkin diagram of $g$.)

This relation was first deduced in the Liouville model, which is
$sl(2)$ Toda theory. A necessary requirement that the Liouville theory
be conformally invariant is that the potential term has  conformal
weight (1,1). One may  obtain the Virasoro algebra from the Lagrangian
by adding a conformal improvement to the stress-energy tensor. This
then ensures that the potential term has the correct weight. Having
obtained the Virasoro algebra for the Liouville theory at coupling constant
$\beta$, there is no other value $\beta'$ for which the potential
term in the classical Liouville theory with coupling constant $\beta'$
has weight (1,1) with respect to the Virasoro algebra in the Liouville
theory with coupling constant $\beta$.

In the quantum theory, however, there are $O(\hbar)$ corrections to the
stress-energy tensor which allow the potential term in the
quantum theory with coupling constant $1/(\beta\hbar)$ to have weight
(1,1) with respect to the Virasoro algebra in the Liouville theory
with coupling constant $\beta$.

This was used by Mansfield \cite{Mans2} to explain a problem that
was present in the interpretation of the
Liouville model at imaginary coupling constant as the Lagrangian
theory of the conformal minimal models; this problem being that the
quantum solutions to the equations of motion only generated the $(1,q)$
or $(p,1)$ conformal primary fields. With the addition of the
potential term of the `dual' theory,  all the conformal primary
fields $(p,q)$ could be obtained.

In the case of
the Liouville theory, the Virasoro algebra is the full
chiral algebra. For Toda theories based on more general simply laced
algebras, one obtains a W-algebra.
The Drinfeld-Sokolov construction for a Hamiltonian reduction based on
the maximal regular embedding of $sl(2)\hookrightarrow g$
allows one to find explicit
expressions for the chiral algebras of classical conformal Toda
theories, and the conserved quantities in classical affine theories,
through the Miura transformation.
One finds that the conserved quantities are polynomial in the
derivatives of the Toda fields $\phi^i$ and $1/\beta$.
When people sought free-field
representations of W-algebras (the chiral algebras of conformal Toda
theories), they were brought to the idea of quantising the Miura
transformation. The expressions found for \wan and \wdn by Fateev
and Luk'yanov \cite{FLuk1,FLuk2}
are exactly the same as the classical expressions, after
normal ordering, with the replacement of the coupling constant $1/\beta$
by $1/\beta - (\hbar\beta)$.
These W-algebras have a representation theory quite analogous to that
of the Virasoro algebra, and similarly the naive quantisation of the
Lagrangian only generates some of the primary fields. However, these
quantum W-algebras also admit an extension to the Lagrangian by the
potential term at the inverted coupling constant \cite{Mans2}, so that
the full set of primary
fields can be obtained from a Lagrangian theory which contains the
potential terms
for both coupling constants $\beta$ and $1/(\hbar\beta)$.

When considering the quantisation of the W-algebra symmetries  of
non-simply laced Toda theories there was much less success,
and a quantum construction of the chiral algebra in the $B_2$ Toda
theory has only
been found very recently \cite{KWat3}.
Although the fields in this quantum construction are  $O(\hbar)$
corrections to the classical expressions, this does not occur simply
as a renormalisation of the coupling constant.
This may be seen from the fact that the extra
potential terms which may be added to the $B_2$ theory are in fact
those of the $C_2$ theory at the dual coupling constant; although
$B_2$ and $C_2$ are isomorphic as Lie algebras, their root systems are
naturally thought of as rotated by $45^\circ$ with respect to each
other.
Thus, equally well the W-algebra fields are $O(1/\hbar)$ corrections
to the $C_2$ theory expressions.
We see that the quantum W-algebras may be
thought of as a simple renormalisation of the coupling constants in
the classical expressions only when the root spaces of $g$ and
$g^\vee$ coincide, which means that $g$ must be simply-laced if it is a Lie
algebra; if $g$ is a super-algebra, the same condition  holds for {\it
WB}$(0,n)$ too, for which many results
analogous to the simply-laced algebras hold \cite{Destetal}.
Thus, the chiral algebra of any Toda theory of
a non-simply laced
algebra will be $O(\hbar)$ corrections to the classical expression for
the chiral algebra, if we replace $\beta$
by $1/(\hbar \beta)$, then
they are $O(\hbar)$ corrections to the classical expressions for
the chiral algebras of the theory based on $g^\vee$.
This result has been very recently proven by Frenkel \cite{fren1}
using homological techniques.

When we come to discuss affine Toda theories we may well ask if
similar statements hold.  It has long been known that the classical
affine Toda theories have an infinite set of conserved quantities
\cite{DSok2}, with spins given by the exponents of the associated
affine Lie algebra.  Some initial work was done on quantum affine Toda
theory by Eguchi and Yang \cite{EYan1}, and by Palla \cite{Pall1}.
Feigin and Frenkel have recently shown the existence of a full set of
quantum conserved currents through the use of homological techniques
\cite{FFre4}. Affine Toda theory splits into two regimes; where the
coupling constants is real, and where it is imaginary. The real
coupling constant regime is more closely associated to perturbed
conformal field theories, but it is for the imaginary coupling
constant behaviour that most progress has been made in understanding
the particle content of the theories
\cite{BCDS1,BCDS2,Destetal}. In this letter we give a  short proof
that the conserved quantities of the affine theories in these regimes
obey the duality already found for the conformal theories. We give
several examples based on our calculations for conformal Toda theories
\cite{KWat3}. We end with a discussion of the implications of this
duality on the mass spectrum for these theories, and the recent
conjectures of Grisaru et al. \cite{grisetal} for the scattering
matrices of some non-simply laced affine theories.

\section{Quantum Toda Theory}
\label{t3}

\def\vecz#1{\vec{#1,\bar z}}
\def\fra#1{{#1}/{a}}
\def\no:#1:{\mathopen\mno #1 \mathclose\mno}
\def\im{\mathop{\rm Im}\nolimits}
\def\dbz{\,d{x^+}\,d{x^-}\,}
\def\dbz{\,d^2x\,}
\def\Ker{\mathop{\rm Ker}\nolimits}

The quantum Toda field theory associated to a finite dimensional
algebra $g$ has the same Lagrangian as the classical theory,
\begin{equation}
  \cL = \half1 \partial_\mu \p \partial^\mu \p - \sum_j m_j
%  \cL = \dpl \p \cdd\dmi \p - \sum_j m_j
  e^{\b\alpha_j\cdd \p}
  \;,
\end{equation}
where $\Sigma(g) = \{\alpha_j\}$ denotes the set of simple roots of $g$,
$m_j$ are rank $g$ arbitrary non-zero constants and $\b$ is
the coupling constant of the theory.
Removing the divergencies by normal ordering introduces a mass scale
\cite{Cole1}. However, the algebraic structure does not depend on
the mass scale, and one can set the constants $m_j$ to any arbitrary
non zero values by a constant shift in $\phi$.

We  work in light-cone coordinates,
$z, \bar z$ given by
\begin{equation}
z = e^{i\sqrt2 x^+}\;,\;\; \bar z = e^{-i\sqrt2 x^-}\;,
\end{equation}
where $0 \leq x \leq \sqrt2 \pi$,
$x^\pm = (t \pm x)/\sqrt2$.
%We find it more convenient to use the variables
We shall also denote $\partial/\partial z$ by $\partial$, and
define new fields $X,H$ by
\begin{eqnarray}
  X(z,\bar z) &=& -i \sqrt2 \p(x^+,x^-) \;, \\
  H(z,\bar z) dz &=& \sqrt2 \dpl\p(x^+,x^-) dx^+ \;,
\end{eqnarray}
and similarly for $\bar H$. Note that $H = i\partial X$.
In the quantum theory the conformal symmetry is
generated by an improved stress-energy tensor of the form
\begin{equation}
  T_{zz} = \half 1 H\cdd H + (a\rho - \rho^\vee/a)\cdot\partial H
  \;,\label{e1}
\end{equation}
where $\rho = \half 1 \sum_{\Delta^+} \alpha$, $\rho^\vee = \half
1\sum_{\Delta^+} \alpha^\vee$, $\hbar$ has been set to 1, and $a =
\b/\sqrt2$ is the coupling constant.
$\Sigma^+$ is the set of positive roots of $g$, and
the dual roots are given by $\alpha^\vee =
2 \alpha / \alpha^2$.
We find that the improved stress energy tensor is
conserved \cite{Mansfield.earlier.toda} and so we can decompose it
into modes as
%\begin{eqnarray}
\eq
  T_{zz} = \sum_n
L_n z^{-n-2} \quad\quad,\qquad\qquad
  T_{\bar z\,\bar z} = \sum_n \bar L_n \bar z^{-n-2} \;,
\en%\end{eqnarray}
where $\bar\partial L_n=0$ by the Heisenberg equations of motion.
We use light-cone commutation relations, which are
of the form
\begin{equation}
  [H^j(z),H^k(z')] = \delta^{jk}\delta'(z-z')
  \;.
\label{eq.hcom}
\end{equation}
We expand the fields $X(z,\bar z)$ and $H(z,\bar z)$ in equal-$x^-$
modes,
\begin{eqnarray}
  X^j(z,\bar z) &=& q^j(\bar z) - i p^j(\bar z) \ell n z +
	\sum_{n\neq 0} iH^j(\bar z)_n z^{-n}/n \\
  H^j(z,\bar z) &=& \sum_n H^j_n(\bar z) z^{-n-1}
\end{eqnarray}
for $x^- < x^+ < x^- + \pi/\sqrt2$. We have also set $H^j_0 \equiv
p^j$.
Similarly we expand in equal-$x^+$ modes for $x^+ < x^- < x^+ + \pi/\sqrt2$.
The commutation relations for the modes are
\eq%\begin{eqnarray}
  {}[H^j(\bar z)_m,H^k(\bar z)_n] = m \d^{jk}\delta_{m+n}
	 \qquad,\qquad\qquad
  {}[q^j(\bar z),p^k(\bar z)] = i\d^{jk}
\en%\end{eqnarray}
We shall in future drop reference to the coordinate $\bar z$ where the
meaning is unambiguous. We have a Hilbert space for each $\bar z$ on
which these modes act, with a vacuum $\vac$.  A field $\psi(z,\bar z)$
satisfies
\begin{equation}
  \psi(z,\bar z)\vac = e^{zL_{-1}} \vecz\psi\;.
\label{eq.tr}
\end{equation}
Thus, we may evaluate the equal-$\bar z$ commutation relations of
fields and states exactly as in standard conformal field
theory \cite{Godd1}, providing we remember that the $\bar z$ dependence
of our states and modes is non-trivial.

If we define $E^j(z,\bar z) dz d\bar z =
\no: e^{ia\a_j\cdd X(z,\bar z)} : 2dx^+ dx^-$ then using standard
conformal field theory normal ordering we obtain
\begin{eqnarray}
  E^j(z,\bar z) &=&
	{\exp\left[\sum_{n>0}a\a_j\cdd H(\bar z)_{-n}z^n/n \right]}
	e^{ia\a_j\cdd q(\bar z)}
	z^{a\a_j\cdd p(\bar z)} \cont\times
  	{\exp\left[\sum_{n>0} -a\a_j\cdd H(\bar z)_n
		z^{-n}/n \right]}
  \;.
\end{eqnarray}
The quantum equations of motion are
\begin{equation}
  \bar\partial H(\bar z)_n =
	-\half 1\sum_j am_j\alpha_j E^j(\bar z)_n,
\end{equation}
where $E^j_n$ are the modes of the vertex operator,
\begin{equation}
  E^j(z,\bar z) = \sum_n E^j(\bar z)_n z^{-n-1}.
\end{equation}
By the uniqueness of vertex operators and the conservation of the
stress-energy tensor, to evaluate the
derivative of a field,
we need only evaluate the derivative of the
state to which it corresponds,
\begin{eqnarray}
\bar\partial \psi(z,\bar z)\vac&=&
	\bar\partial e^{zL_{-1}} \vecz\psi \\
	&=& e^{zL_{-1}} \bar\partial \vecz\psi
\end{eqnarray}
Thus, we may easily evaluate the derivative of any polynomial in the
field $H$.
\begin{equation}
  \bar\partial \vecz W = -\half1 \left[\sum_{n,j} am_j
	E^j(\bar z)_{-n}\alpha_j \cdd
	\frac{\delta}{\delta H(\bar z)_{-n}} \right]
	\vecz
W
\;.
\end{equation}
Conversely, since $E^j(\bar z)_0\vac=0$,
and $[L_n, E^j(\bar z)_0] =0$, we may check directly from the
commutation relations
that
\begin{eqnarray}
  {}[E^j(\bar z)_0,W(0,\bar z)]\vac &=&
	E^j(\bar z)_0 \vecz W \\
  &=& -\left[ \sum_n a E^j(\bar z)_{-n}\alpha_j \cdd
	\frac{\delta}{\delta H(\bar z)_{-n}} \right]
	\vecz W
\;.
\end{eqnarray}
And thus a field $W[H]$
is chiral if and only if it commutes with all
the operators $Q^j = E^j(\bar z)_0$. These operators $Q^j$ are
commonly called the screening charges.
The Virasoro algebra, generated by the modes of the stress energy
tensor, commutes with all the screening charges $Q^i$ and is given by
the field corresponding to the state
\eq
\vec L = \left[
\half 1(H_{-1}\cdd H_{-1}) + (a\rho - \rho^\vee /a)\cdd H_{-2} \right]\vac
\;.
\label{eq.lstat}
\en

\section{Proof of duality of chiral algebras}

There is another set of screening charges
\begin{equation}
  Q^{i\vee} = \int dz \no:e^{-i/a \alpha_i^\vee\cdd X}:
\end{equation}
which
corresponds to the Toda theory of $g^\vee$, at coupling constant
$a^\vee = -1/a$.
We find that the W-algebra given by the commutant
of the screening charges $Q^i$ of $g$ also commutes with the
screening charges $Q^{i\vee}$ of $g^\vee$.
This has been proven in full generality by Frenkel \cite{fren1}. For
our extension to the affine case we need only consider the regime
$a^2$ negative. We think it useful to present here a version of the
proof of the duality of the chiral algebra in this regime.
%\subsection{Proof of duality of chiral algebras}

We consider each simple root in turn.
We are seeking the commutant of $Q^i$ in the space of fields spanned
by polynomials
in $H^j$ and their derivatives. Since we know the relation between
states and fields (\ref{eq.tr}) is unique,
$[Q^i,\psi(z)]=0$ is equivalent to $Q^i \vec\psi=0$.
Since the modes $H^j_m$ have the
commutation relations (\ref{eq.hcom}), we can decompose the space of
modes at each level into the modes
$K^i_m = \a_i/|\a_i|\cdd H_m$ and an $n-1$
dimensional space of modes which commute with the modes $\a_i\cdd
H_m$. For each $i$ the Fock space of the modes $H^j_m$ splits into the
tensor product of the Fock spaces of the modes $K^i_m$ and the
orthogonal modes,
\begin{equation}
\cH = \cH_K \otimes \cH^\perp.
\end{equation}
Since the action of the operator $Q^i$ can be expressed entirely
in terms of the modes $K^i_m$, we see that
\begin{equation}
\Ker_\cH(Q^i) = \cH^\perp\otimes\Ker_{\cH_K}(Q^i).
\end{equation}
Thus we only need consider this simpler space $\Ker_{\cH_K}(Q^i)$.
We have chosen the modes $K^i_m$ to have free field commutation
relations,
\begin{equation}
{}~[K^i_m,K^i_n] = m\delta_{m+n,0}
\end{equation}
and the operator
$Q^i$ is expressed as $\int dz \exp(ia \int K(z)dz)$.
We know at least one state in $\cH_K$ which is not in
$\Ker_{\cH_K}(Q^i)$, this being $K^i_{-1}\vac$, since
\begin{equation}
Q^i K^i_{-1}\vac = a \vec{ a\a_i}.
\end{equation}
We also know that the Virasoro algebra $L^i_m$ given by
\begin{equation}
L^i(z) = \half 1\no:K^i(z)^2: + \left[
	a|\a_i|/2 - 1/(a|\a_i|)\right]
	 \partial K^i(z)
\end{equation}
commutes with $Q^i$. Thus the space
$\Ker_{\cH_K}(Q^i)$ is a representation of the Virasoro algebra $L^i$.
Similarly, the states produced by the action of $Q^i$ on $\cH_K$, which
we denote by $\im_{\cH_K}(Q^i)$ form a representation of
the Virasoro algebra. Since $Q^i K^i_{-1}\vac \ne 0$, we know that
this representation has at least one state with $h=1$, and so it must
be a highest weight representation with $h=1$.
Using the Kac determinant formula \cite{GOli1}, we know that if $a^2$
is negative, then the $h=1$
representation is irreducible. This representation thus has character
\eq
\chi_{h=1}(q) =  q \prod_{n=1}^\infty (1-q^n)^{-1}
\en
which must be a lower bound on the character of $\im_{\cH_K}(Q^i)$.
Consequently there is an upper bound on the character of the space
$\Ker_{\cH_K}(Q^i)$, namely
\eq
(1-q)\prod_{n=1}^\infty (1-q^n)^{-1} \;.
\label{eq.ch}
\en
$\Ker_{\cH_K}(Q^i)$ contains one copy of a highest weight
representation with $h=0$. However, we know the character of the $h=0$
representation, and it is exactly given by (\ref{eq.ch}).
Since (\ref{eq.ch}) was an upper bound on the character of
$\Ker_{\cH_K}(Q^i)$
we can deduce that
%$\tilde\chi=0$ and that
$\Ker_{\cH_K}(Q^i)$ is an irreducible
highest weight representation with $h=0$.
Thus all the states in $\Ker_{\cH_K}(Q^i)$ are given by the
action of creation modes of the algebra $L^i$ acting on the vacuum.

So, we can decompose the Fock Space $\cH$ as
\eq
\cH = \cH^{i\perp} \otimes ( V_0 \oplus V_1 ) \;,
\label{eq.deco}
\en
where $V_0$ is given by the action of all creation modes of $L^i$ on
$\vac$, and $V_1$ by the action of all creation modes of $L^i$ on
$K^i_{-1}\vac$.

Thus we can write a generic state $\vec\psi$ in $\cH$ as
\eq
\vec\psi = \sum_{j\in I} a_j \vec{p_j} \otimes \left(
	q_j[ L^i] \vac + r_j[L^i] K^i_{-1}\vac \right)
\en
where $\vec{p_j}$ is an arbitrary state in $\cH^{i\perp}$ and $q_j,r_j$ are
polynomials in the lowering modes of $L^i$.
If we consider the condition that this state defines a field
which commutes with $Q^i$ we obtain
\eqq
Q^i\vec\psi  &=& a \sum_{j\in I} a_j \vec{p_j} \otimes r_j[L^i]
\vec{a\alpha_i} \nonumber\\
	&=& 0  \;.
\label{eq.qpo}
\enn
We can now ask that the same state defines a field which commutes with the
dual screening charge $Q^{i\vee}$.
The $Q^{i\vee}$ have the property that $[ L_i, Q^{i\vee}]=0$ and so
\eqq
Q^{i\vee}\vec\psi &=& (-1/a) \sum_{j\in I} a_j \vec{p_j} \otimes r_j[L^i]
\vec{-1/a \alpha^\vee_i} \nonumber\\
	&=& 0  \;.
\label{eq.qpt}
\enn
These are the same equations for the state $\vec{p_j}$ and the
polynomials $r_j$, and so any state which commutes with $Q^i$
also commutes with $Q^{i\vee}$,
\begin{equation}
\Ker_{\cH_K}(Q^i) = \Ker_{\cH_K}(Q^{i\vee}).
\label{eq.p1}
\end{equation}
The chiral algebra of the Toda theory for $a^2$ negative is given as
\eq
{\cal A} = \bigcap_i \Ker_{\cH_K}(Q^i)
\en
and so all the fields in $\cA$ also automatically commute with the
operators $Q^{i\vee}$. Thus if we denote the chiral algebra of the
Toda theory based on $g$ with coupling constant $a$ by $\cA(g,a)$,
then we have
\begin{equation}\label{eq:algdual}
\cA(g,a) = \cA(g^\vee, -1/a).
\label{eq.ag}
\end{equation}
Note that in (\ref{eq:algdual}) the simple roots of $g^\vee$ are
$2\a/\a^2$, where $\a$ are the simple roots of $g$.

For $a^2$ positive, rational, there are subtleties in that the
Virasoro representation $\Ker(Q^i)$ is not irreducible and so the
decomposition (\ref{eq.deco}) is invalid.

\section{Affine algebras and examples}

\def\Aff{\mathop{\it Aff}\nolimits}
\def\Af{\mathop{\it aff}\nolimits}

The classical affine Toda field theories are integrable field theories
of scalar fields related to affine Lie algebras. If $\Sigma=
\{\alpha_i\}_{i=0,\ldots,l}$ is the set of simple roots of an affine
Lie algebra $\hat g$ of rank $l$,
then one may define the Lagrangian for a theory of $l$ scalar fields
as
\begin{equation}
  \cL = \half1 \partial_\mu \p \partial^\mu \p - \sum_{i=0}^l m_i
  e^{\b\a_i\cdd\p}
  \;.
\end{equation}
We shall again consider this theory on the light-cone, and use the
conventions of the previous section.
When one considers normal ordering the Lagrangian, one must now be
more careful. However, one can show that a change in the mass scale
can again be compensated for by a shift in $\phi$, in the sense that
requiring $\langle\phi\rangle = 0$ implies that the $m_j = M n_j$, where
$\sum n_j \a_j = 0$, and $M$ is a constant which depends on
the mass scale. Thus the form of the Lagrangian and in particular the
value of $\b$ are independent of the regularisation scheme
\cite{deveganew}.
For the conformal Toda theories, the conserved quantities were local
fields, giving a W-algebra as the symmetry algebra. However, for
affine theory this is not the case.
Instead of all the modes of a local field commuting with the
Hamiltonian, only the integrals of fields will commute.
The conserved quantities of this
integrable system are given by
\begin{equation}
\cI_a = \int W_a dz,
\end{equation}
where $\cI_a$ and $W_a$ satisfy
\begin{equation}
{}~[\cH_0,\cI_a]=0 \;\;,\; [\cH_0,W_a] = \partial V_a \;.
\label{eq.w1}
\end{equation}
Here
$\cH_0$ is the Hamiltonian of the theory and $V_a$ are arbitrary local
fields.
A field $W$ satisfies (\ref{eq.w1}) if and only if
\eq
\cH_0 \vec {W_a} = L_{-1} \vec{V_a} \;,
\en
where $L_{-1} = \int dz \half 1 \no:H\cdot H:$.
The quantum Hamiltonian is
\begin{equation}
\cH_0 = \int dz \sum_{j=0}^l m_j\no: exp( i a \a_j \cdd X): \;,
\label{eq.h2}
\end{equation}
and so, by the algebraic independence of the exponential terms, we see
that the conserved quantities  $\{\cI_a\}$ must commute with each of
the terms in the Hamiltonian individually.
The conserved quantity is called trivial if $W_a = \partial
Y_a$ where $Y_a$ is a local field.
Since $L_{-1}$ commutes with $\cH_0$ we have for a trivial conserved
quantity $\vec{W_a} = L_{-1} \vec{Y_a}$.
Let us denote by $\Aff(g,a,j)$ the space of states $\vec W$ such that
\eq
Q^j \vec W = L_{-1} \vec V \;,
\label{eq.w2}
\en
where $Q^j = \int dz \no:exp(ia\a_j\cdd X):$.
Then the total space of states satisfying (\ref{eq.w1}), which we
denote by $\Aff(g,a)$, is given by
\eq
\Aff(g,a) = \bigcap_{i=0}^l \Aff(g,a,i) \Big/ L_{-1} \cH \;.
\en
Feigin and Frenkel have recently shown that the space of states so
defined has
the same dimensionality as in the classical case \cite{FFre4}. There is one
conserved quantity of spin $e_a$ for each exponent of the affine
algebra $\hat g$, corresponding to a state $\vec{W_a}$ of conformal
weight $e_a+1$. We shall now show that in the regime $a^2$ negative,
the conserved quantities of the affine theory $\hat g$ are also
conserved quantities of the theory $\hat g^\vee$ with inverted
coupling constant.

We shall again consider each simple root in turn and introduce the
operators $K_m$ associated to the
particular simple root under consideration.
If we consider the decomposition (\ref{eq.deco}) of the space $\cH$,
then we may again take an arbitrary state in $\cH$ to be
\eq
\vec\psi = \sum_{j \in I} a_j \vec{p_j} \otimes \left(
	q_j[ L^i] \vac + r_j[L^i] K^i_{-1}\vac \right),
\en
where $\vec{p_j}$ is an arbitrary state in $\cH^{i\perp}$ and $q_j,r_j$ are
polynomials in the creation modes of $L^i$.
If we consider the condition that this state defines a field
whose integral commutes with $Q^i$ we obtain
\eqq
Q^i\vec\psi  &=& a \sum_{j\in I} a_j \vec{p_j} \otimes  r_j[L^i]
\vec{a\alpha_i} \nonumber\\
	&=& L_{-1} \sum_{j \in I} a_j \vec{p_j} \otimes (
	q_j[ L^i] \vac + r_j[L^i] K^i_{-1}\vac ) \;.
\label{eq.qpo2}
\enn
If we ask that the same state defines a field which commutes with the
dual screening charge $Q^{i\vee}$, we obtain
\eqq
Q^{i\vee}\vec\psi &=& (-1/a) \sum_{j \in I} a_j \vec{p_j} \otimes  r_j[L^i]
\vec{-1/a \alpha^\vee_i} \nonumber\\
 	&=& L_{-1} \sum_{j \in I} a_j \vec{p_j} \otimes (
	q_j[ L^i] \vac + r_j[L^i] K^i_{-1}\vac )  \;.
\label{eq.qpt2}
\enn
These are the same equations for the states $\vec{p_j}$ and the
polynomials $r_j,q_j$ and so the integral of any field which commutes
with $Q^i$ also commutes with $Q^{i\vee}$, and so
\begin{equation}
\Aff(g,a,i) = \Aff(g^\vee, -1/a,i) .
\end{equation}
Since $L_{-1}$ is independent of $g$ and $a$ we have
\begin{equation}
\Aff(g,a) = \Aff(g^\vee, -1/a)
\label{eq.pp1}
\en
and the two affine Toda theories have the same conserved quantities.
\section{Examples}

\def\st{\mathop{\rm-}}

The classification of affine Dynkin diagrams may be found in
\cite{Kac1}; we shall adopt this notation.
Each affine diagram with $(l+1)$ vertices defines a theory of $l$ scalar
fields.
We have investigated the theories $C_2^{(1)}, D_3^{(2)},
G_2^{(1)}, D_4^{(3)}, B_3^{(1)}, A_5^{(2)}, C_3^{(1)}, D_5^{(2)},
A_4^{(2)}, A_6^{(2)}$.
With the exception of all but the last two, these can be split into
dual pairs, for which the Dynkin diagrams are
as follows,
\vskip .2cm
\eq
\begin{array}{cl@{\qquad\qquad}cl}
\multicolumn{2}{c}{\hat g\qquad\qquad}& \multicolumn{2}{c}{\hat g{}^\vee} \\
C_2^{(1)} & \Gfiv &
D_3^{(2)} & \Gsix \\
G_2^{(1)} &
\begin{picture}(30,10)(-5,0)
\adynking2
\end{picture} &
D_4^{(3)} &
\begin{picture}(30,10)(-5,0)
\adynkind4
\end{picture} \\
B_3^{(1)} & \Gone &
A_5^{(2)} & \Gtwo \\
C_3^{(1)} & \Gthr &
D_5^{(2)} & \Gfou
\end{array}
\nonumber
\en
\vskip .5cm

For the case $A_4^{(2)}$ and $A_6^{(2)}$, these are self dual with
Dynkin diagrams
\vskip .2cm
\eq
\begin{array}{cl@{\qquad\qquad}cl}
\multicolumn{2}{c}{\hat g\qquad\qquad}& \multicolumn{2}{c}{\hat g{}^\vee} \\
A_4^{(2)} & \Gsev &
A_4^{(2)} & \Geig \\
A_6^{(2)} & \Gnin &
A_6^{(2)} & \Gten
\end{array}
\nonumber
\en
\vskip .5cm
We have searched for conserved quantities of the affine Toda theories
based on these Dynkin diagrams up to spin 5 for the rank three cases, and
to spin 7 for the rank two cases.
This was done using the algebraic manipulation language REDUCE to
evaluate the action of $Q^i$ and $Q^{i\vee}$.
If we consider fields in the chiral algebra $\cA(g,a)$, and equation
(\ref{eq.w2}) for the screening charge corresponding
to the additional root, we only need consider states in $\cH$ which
are formed from the action of lexicographically ordered modes of the
fields in the chiral algebras. We can then compare this with an
arbitrary states of the form $L_{-1}\vec V$.

\subsection{$B_2$}

Here explicit expressions for the chiral algebra are known
\cite{KWat3}. One finds unique states $\vec W$ such that
$\cH_0 \vec W = L_{-1}\vec V$ for the spin of $W$ equal to $2,4$ and $6$.

We take the simple roots of $B_2$ to be
\eq
( 0, 1 ), \qquad ( 1,-1)
\en
If we consider the conformal theory Toda theory based on $B_2$ then we can
find fields which commute with the screening charges for the simple
roots of the algebra. Starting with an arbitrary state at each level
we implemented these constraints, and up to conformal weight 6, we
found two independent fields which satisfied them. One was the
Virasoro algebra generator, which corresponded to the state
(\ref{eq.lstat}).
The other was a spin 4 conformal primary field corresponding to
the state $W_{-4}\vac$, which we have described in our paper \cite{KWat1}.
The field has an extremely lengthy expression in terms of free fields
which is an $O(1/a)$
perturbation of the classical expression,
and we choose to normalise the state $W_{-4}\vac$ by
\begin{eqnarray}
\langle 0 | W_4 W_{-4} \vac &=&
 - 72 (3a - 2/a)(a - 3/a) (5a - 6/a)(3a - 5/a) (5a - 8/a) \cont\times
	(4a - 5/a) (7a - 6/a)(3a - 7/a)
	\Big/ (75a^2 - 226 + 150/a^2)
%
%(7a^2 - 6)(5a^2 - 6)(5a^2 - 8)(4a^2 - 5)
%(3a^2 - 2)(3a^2 - 5)(3a^2 - 7)(a^2 - 3)
%
%(75a^4 - 226a^2 + 150)a^6
%
\end{eqnarray}
We also found  of course the Virasoro descendants of $L_{-2}\vac$ and
of $W_{-4}\vac$,
and we give the whole list in Table \ref{tab.wfields}.

Having found these fields, we may search for conserved quantities
for the quantum affine theories related to $B_2$. These affine
theories are
$C_2^{(1)}$ and $D_3^{(2)}$, which form a pair of dual algebras.
We can also consider $A_4^{(2)}$ to be an extension of $B_2$ in two
different ways; this is a self dual algebra.
The additional roots are as
\begin{Table}[h]
\caption{Affine extensions of $B_2$}
\renewcommand{\arraystretch}{1.2}
\begin{center}
\begin{tabular}[c]{ccccc}
algebra & $C_2^{(1)} $  & $ D_3^{(2)} $ & $A_4^{(2)}$ (i) &
$A_4^{(2)}$ (ii) \\
root & $(-1,-1) $		& $(-1,0)$	& $(-1/2,-1/2)$ & $(-2,0)$
\end{tabular}
\end{center}
\end{Table}

The conserved quantities for these affine theories
are of the form
\eq
\cI = \int \Phi(z) d\!z
\label{eq.conform}
\en
where $\Phi(z)$ is some field which is polynomial in the derivatives
of the Toda fields. This charge $\cI$ must certainly commute with the
Hamiltonian of the conformal theory, which comprises the screening
charges for the conformal theory, and so we can take $\Phi(z)$ to be
some field in the W-algebra of the conformal Toda theory.  With
$\Phi(z)$ of this form, we now need only check that the integral $\cI$
commutes with the screening charges corresponding to the extra term in
the potential of the affine theory.

We systematically searched for fields $\Phi(z)$ of this form up to
spin $8$. We found only three such fields, although there are many more
in the W-algebra up to conformal spin 8 which we list in Table
\ref{tab.wfields} below.
To find conserved quantities, we only need consider fields $\Phi(z)$
which cannot be expressed as the $\partial / \partial z$ derivative of
another local field, or for which the state $L_{1}\vec\Phi = 0$.
Thus, when we look for conserved quantities, we need only look at
fields $\Phi$ of conformal spins 2, 4 and 6.
We find, as expected, that the momentum
\eq
\int L(z) d\!z
\en
is a conserved quantity.
There are no conserved quantities at spin 2, or suitable $\Phi$ at
spin 3.
At $\Phi$ of spin 4, for a conserved quantity of spin 3, we found a conserved
charge for each of the
affine theories. If we write the
field $\Phi$ as
\eq
\Phi_{-4}\vac = W_{-4}\vac + x L_{-2}L_{-2} \vac
\en
then we obtain the conserved quantities for the choices of $x$ below.
\begin{Table}[h]
\begin{center}
\caption{Spin 3 charges}
\vskip .2cm
\begin{tabular}[c]{|c|c|c|c|c|}
\hline \emptycol{5} \\
  & $C_2^{(1)}$ 		& $D_3^{(2)}$ & $A_4^{(2)}$ (i) &
$A_4^{(2)} $ (ii) \\
\sline{5}
$x
$&$
{{18 (3 a^2 -2)^2 ( 3a^2 -7)}
	\over{(3a^2-4)(75 a^4 - 226 a^2 + 150)}}
$&$
{{ - 18 (a^2 - 3)^2 ( 7a^2 - 6)}
	\over{(2a^2-3)(75 a^4 - 226 a^2 + 150)}}
$&$
{{
18(4 a^2 - 5)(5 a^2 - 22)
}\over{
5 (75 a^4 - 226 a^2 + 150)
}}
$&$
-
{{
18(5 a^2 - 8)(11 a^2 - 5)
}\over{
5(75 a^4 - 226 a^2 + 150)
}}
$
\\ \emptycol{5}\\
\hline
\end{tabular}
\end{center}
\end{Table}
%
%	table of W-algebra fields moved here because of spacing
%
\begin{table}[ht]
\begin{center}
\caption{W-algebra states for $B_2$}
\label{tab.wfields}
\vskip .2cm\renewcommand{\arraystretch}{1.2}
\begin{tabular}[c]{|c|c|}
\hline\emptycol{2}\\
  spin 		& state \\
\sline{2}
 $2$		& $L_{-2}\vac $\\
\sline{2}
 $3$		& $L_{-3}\vac $\\
\sline{2}
 $4$		& $L_{-4}\vac\,,\, W_{-4}\vac\,,\, L_{-2}L_{-2} \vac $ \\
\sline{2}
 $5$		& $L_{-5}\vac\,,\, W_{-5}\vac\,,\, L_{-3}L_{-2} \vac $ \\
\sline{2}
 $6$		& $L_{-6}\vac\,,\, W_{-6}\vac\,,\, L_{-3}L_{-3}\vac $ \\
		& $L_{-4}L_{-2}\vac\,,\, L_{-2}W_{-4}\vac\,,\,
			L_{-2}L_{-2}L_{-2} \vac $ \\
\sline{2}
		& $L_{-7}\vac\,,\, W_{-7}\vac\,,\, L_{-4}L_{-3}\vac $ \\
 $7$		& $L_{-5}L_{-2}\vac\,,\, L_{-2}W_{-5}\vac\,,\,
			L_{-3}W_{-4}\vac $ \\
		& $L_{-3}L_{-2}L_{-2} \vac $ \\
\sline{2}
		& $L_{-8}\vac\,,\, W_{-8}\vac\,,\, L_{-4}L_{-4}\vac $ \\
 		& $L_{-5}L_{-3}\vac\,,\, L_{-6}L_{-2}\vac\,,\,
			L_{-2}W_{-6}\vac $ \\
 $8$		& $L_{-3}W_{-5}\vac\,,\, L_{-4}W_{-4}\vac\,,\,
			W_{-4}W_{-4}\vac $ \\
		& $L_{-2}L_{-2}W_{-4}\vac\,,\, L_{-3}L_{-3}L_{-2}\vac $ \\
		& $L_{-4}L_{-2}L_{-2}\vac\,,\,
			L_{-2}L_{-2}L_{-2}L_{-2}\vac $ \\
\emptycol{2}\\\hline
\end{tabular}
\end{center}
\end{table}

At $\Phi$ of spin 6, for a conserved quantity of spin 5, we found a
unique conserved charge for the affine theories $C_2^{(1)}, D_3^{(2)}$.
If we write
\eq
\vec\Phi = ( L_{-2}W_{-4} + x L_{-2}L_{-2}L_{-2} + y L_{-4}L_{-2})
\vac
\en
then we get:

\begin{Table}[h]
\begin{center}
\caption{Spin 5 charges}
\renewcommand{\arraystretch}{1}
\vskip .2cm
\begin{tabular}{|c|c|c|}
\hline \emptycol{3}\\
  & $C_2^{(1)}$ 		& $D_3^{(2)}$ \\
\sline{3}
$x$ &
$
{{
2(a^2 - 3)(93 a^4 - 160 a^2 + 84)
}\over{
(3a^2 -4)(75 a^4 - 226 a^2 + 150)
}}
$&$
{{
-2(3a^2 - 2)(21 a^4 - 80 a^2 + 93)
}\over{
(2a^2 -3)(75 a^4 - 226 a^2 + 150)
}}
$\\ \emptycol{3}\\
$y$ &
$
-{{
6(a^2 -3)(120 a^8 - 493 a^6 + 720 a^4 - 464 a^2 + 120)
}\over{
a^2(3a^2 -4)(75 a^4 - 226 a^2 + 150)
}}
$
&
$
{{6(3a^2 - 2)(15 a^8 - 116 a^6 + 360 a^4 - 493 a^2 + 240)
}\over{
a^2(2a^2 -3)(75 a^4 - 226 a^2 + 150)
}}
$\\ \emptycol{3}\\
\hline
\end{tabular}
\end{center}
\end{Table}

We also found a conserved quantity of spin 7 for each of the
affine theories. This charge corresponds to a field $\Phi$ of spin 8,
which can be written in terms of the W-algebra fields as
\begin{eqnarray}
\vec\Phi &=& W_{-4}W_{-4}\vac + x L_{-2}L_{-2}W_{-4}\vac +
	y L_{-4}W_{-4}\vac + z L_{-2}L_{-2}L_{-2}L_{-2}\vac \cont+
	u L_{-4}L_{-2}L_{-2}\vac + v L_{-6}L_{-2}\vac
\end{eqnarray}
The conserved quantities are then obtained for the choices:
\begin{Table}
\begin{center}
\caption{Spin 7 charges}
\label{tab.b2_8}
\vskip .2cm
\begin{tabular}{|c|c|}
\hline\emptycol{2}\\
& $C_2^{(1)}$ \\\sline{2}$
x$&$
{\frac {84\, (33\,a^{6}-107\,a^{4}+84\,a^{2}-12
 )}{ (3\,a^{2}-4 ) (75\,a^{4}-226\,a^{2}+150
 )}}
$\\\emptycol{2}\\$
y$&$
-{28 \,
(180\,a^{10}-1299\,a^{8}+3257\,a^{6}-3444\,a^{4}+1564\,a^{2}-240) \over
a^{2}\,  (3\,a^{2}-4 )
 (75\,a^{4}-226\,a^{2}+150 )
}
$\\\emptycol{2}\\$
z$&$
{12 \,(47583\,a^{12}-427614\,a^{10}+1478251\,a^{8}-2557008\,a^{6}+2366936\,a^
{4}-1108752\,a^{2}+201168) \over
 (3\,a^{2}-4 )^{2}
 (75\,a^{4}-226\,a^{2}+150 )^{2}}
$\\\emptycol{2}\\$
u$&$
-{\frac {72\, (40770\,a^{16}-471519\,a^{14}+
2246016\,a^{12}-5799543\,a^{10}+8966752\,a^{8}-8598320\,a^{6}+5032712
\,a^{4}-1644112\,a^{2}+227040 )}{a^{2} (3\,a^{2}-4 )^
{2} (75\,a^{4}-226\,a^{2}+150 )^{2}}}
$\\\emptycol{2}\\$
v$&$\begin{array}{l}\scriptstyle
72\,(1633500\,a^{20}-23455350\,a^{18}+143927265\,a^{16}-498307912\,a^{14}+
1081746337\,a^{12}-1545601596\,a^{10}
\\\qquad\scriptstyle+
1480281944\,a^{8}-944031488\,a^{
6}+385622640\,a^{4}-91324800\,a^{2}+9504000)
\\ \big/\scriptstyle\left(
5\,a^{4} (3\,a^{2}-4)^{2} (75\,a^{4}-226\,a^{2}+150 )^{2}\right)
    \end{array}
$\\\emptycol{2}\\\hline
\hline\emptycol{2}\\
& $D_3^{(2)}$ \\\sline{2}$
x$&$
-{84\,(3\,a^{6}-42\,a^{4}+107\,a^{2}-66) \over
 (2\,a^{2}-3 )
 (75\,a^{4}-226\,a^{2}+150 )
}
$\\\emptycol{2}\\$
y$&$
{\frac {28\, (30\,a^{10}-391\,a^{8}+1722\,a^{6}-3257
\,a^{4}+2598\,a^{2}-720 )}{a^{2} (2\,a^{2}-3 )(75\,a^{4}-226\,a^{2}+150
 ) }}
$\\\emptycol{2}\\$
z$&$
{\frac {12\, (12573\,a^{12}-138594\,a^{10}+
591734\,a^{8}-1278504\,a^{6}+1478251\,a^{4}-855228\,a^{2}+190332
 )}{ (2\,a^{2}-3 )^{2} (75\,a^{4}-226\,a^{2}+150
 )^{2}}}
$\\\emptycol{2}\\$
u$&$
-{72\,(
7095\,a^{16}-102757\,a^{14}+629089\,a^{12}-2149580\,a^{10}+4483376\,a^
{8}-5799543\,a^{6}+4492032\,a^{4}-1886076\,a^{2}+326160) \over
a^2\, (2\,a^{2}-3 ) (75\,a^{4}-226\,a^{2}+150 )}
$\\\emptycol{2}\\$
v$&$\begin{array}{l}\scriptstyle
 72\,(148500\,a^{20}-2853900\,a^{18}+24101415\,a^{16}-118003936\,a^{14}+
370070486\,a^{12}-772800798\,a^{10}
\\\qquad\scriptstyle+
1081746337\,a^{8}-996615824\,a^{6}
+575709060\,a^{4}-187642800\,a^{2}+26136000)
\\ \big/\scriptstyle\left(
5\,a^{4} (2\,a^{2}-3 )^{2} (75\,a^{4}-226\,a^{2}+150 )^{2}\right)
    \end{array}
$\\\emptycol{2}\\\hline
\end{tabular}
\end{center}
\end{Table}

\begin{Table}
\begin{center}
Table \ref{tab.b2_8}: Spin 7 charges (continued) \\
\vskip .2cm
\begin{tabular}{|c|c|}
\hline\emptycol{2}\\
& $A_4^{(2)}$(i) \\\sline{2}$
x$&$
{84\,(30\,a^{4}-113\,a^{2}+90) \over
5\,\left (75\,a^{4}-226\,a^{2}+150\right )}
$\\\emptycol{2}\\$
y$&$
-{42\,(75\,a^{8}-700\,a^{6}+2332\,a^{4}-2960\,a^{2}+1200) \over
5\,a^2\left (75\,a^{4}-226\,a^{2}+150\right )}
$\\\emptycol{2}\\$
z$&$
{108\,(4\,a^{2}-5)(5\,a^{2}-22)(360\,a^{4}-1243\,a^{2}+930) \over
25\,\left (75\,a^{4}-226\,a^{2}+150\right )^2}
$\\\emptycol{2}\\$
u$&$
-{36\,(82125\,a^{12}-1019625\,a^{10}+5013330\,a^{8}-12708296\,a^{6}+17503220
\,a^{4}-12371400\,a^{2}+3492000) \over
25\,a^2\left (75\,a^{4}-226\,a^{2}+150\right )^2}
$\\\emptycol{2}\\$
v$&$\begin{array}{l}\scriptstyle
72\,(
928125\,a^{16}-16155000\,a^{14}+119232825\,a^{12}-485268080\,a^{10}+
1183204611\,a^{8}
\\\qquad\scriptstyle-
1764835120\,a^{6}+1572362700\,a^{4}-766170000\,a^{2}
+156600000)
\\ \big/\scriptstyle\left(
125\,a^4(75\,a^{4}-226\,a^{2}+150)^2 \right)
    \end{array}
$\\\emptycol{2}\\\hline
\hline\emptycol{2}\\
& $A_4^{(2)}$(ii) \\\sline{2}$
x$&$
-{84\,(45\,a^{4}-113\,a^{2}+60) \over
5\,\left (75\,a^{4}-226\,a^{2}+150\right )}
$\\\emptycol{2}\\$
y$&$
{168\,(75\,a^{8}-370\,a^{6}+583\,a^{4}-350\,a^{2}+75) \over
5\,a^2\left (75\,a^{4}-226\,a^{2}+150\right )}
$\\\emptycol{2}\\$
z$&$
{108\,(5\,a^{2}-8)(11\,a^{2}-5)(465\,a^{4}-1243\,a^{2}+720) \over
25\,\left (75\,a^{4}-226\,a^{2}+150\right )^{2}}
$\\\emptycol{2}\\$
u$&$
-{72\,(
218250\,a^{12}-1546425\,a^{10}+4375805\,a^{8}-6354148\,a^{6}+5013330\,
a^{4}-2039250\,a^{2}+328500) \over
25\,a^2\left (75\,a^{4}-226\,a^{2}+150\right )^{2}}
$\\\emptycol{2}\\$
v$&$\begin{array}{l}\scriptstyle
72\,(
9787500\,a^{16}-95771250\,a^{14}+393090675\,a^{12}-882417560\,a^{10}+
1183204611\,a^{8}
\\\qquad\scriptstyle-
970536160\,a^{6}+476931300\,a^{4}-129240000\,a^{2}+
14850000)
\\ \big/\scriptstyle\left(
125\,a^4(75\,a^{4}-226\,a^{2}+150)^2 \right)
    \end{array}
$\\\emptycol{2}\\\hline
\end{tabular}
\end{center}
\end{Table}

\noindent We have checked explicitly that these conserved quantities
obey the required duality properties, that is they commute with
the terms appearing in the affine Toda Lagrangian for the dual algebra
with $a \to -1/a$.
$C_2^{(1)}$ and $D_3^{(2)}$ are dual algebras, as are
$A_4^{(2)}\rm{(i)}$ and $A_4^{(2)}\rm{(ii)}$. However, the
normalisations of the roots differ from that in (\ref{eq:algdual}), so
that e.g. $x$ of $C_2^{(1)}$ transforms into $x$ of $D_3^{(2)}$ under
$a\longrightarrow -\sqrt2/a$ up to a change of sign in $W$.

The spins of the conserved quantities we have found agree with the
spins of the classical conserved quantities, namely that the spins are
equal to the exponents of the affine Lie algebra (see e.g. Table E,
page 216 \cite{Kac1}).
\begin{Table}[ht]
\begin{center}
\caption{}\renewcommand{\arraystretch}{1.2}
\vskip .2cm
\begin{tabular}[c]{|c|c|}
\hline\emptycol{2}\\
algebra & exponents \\ \sline{2}
$C_2^{(1)}$ & $1 \bmod 2$ \\
$D_3^{(1)}$ & $1 \bmod 2$ \\
$A_4^{(2)}$ & $1,3,7,9 \bmod 10$ \\
\emptycol{2}\\\hline
\end{tabular}
\end{center}
\end{Table}

\subsection{$B_3$}

%\input wbc3.r.tex
% these are the same expressions as before for the variables x.
\def\topspace{\vphantom{\vrule height 3ex depth 0pt}}
\def\bottomspace{\vphantom{\vrule height 0pt depth 2ex}}
\def\zzz{}

We take the simple roots of $B_3$ to be
\eq
(0,0,1), \qquad ( -1,1,0), \qquad  (1,0,-1) \;.
\en
If we consider the conformal theory Toda based on $B_3$ then we can
find fields which commute with the screening charges for the simple
roots of the algebra. Starting with an arbitrary state at each level
we implemented these constraints, and up to conformal weight 6, we
found three independent fields which satisfied them. One was the
Virasoro algebra generator, which corresponded to the state
(\ref{eq.lstat}),
and the other two were a spin 4 and a spin 6 field, corresponding to
states
\eq
W_{-4}\vac \;,\; V_{-6}\vac \;,
\en
respectively.
These are extremely lengthy expressions which are $O(1/a)$
perturbations of the classical expression.
We have used the normalisation
\begin{eqnarray}
  \cev{0} W_4 W_{-4} \vac &=& - 12 (3 a - 5/a) (4 a - 3/a) (5 a  - 7/a)
	(5 a  - 9/a) (6 a  - 7/a) (7 a  - 8/a) \cont\times
	(7 a - 10/a) (9 a  - 8/a) \Big/  (525 a^2  - 1357  + 840/a^2) \\
  \cev{0} V_6 V_{-6} \vac &=& - 112 (a - 3/a) (3 a - 4/a) (3 a - 5/a)
	(4 a - 3/a) (5 a - 3/a) \cont\times
	(5 a - 6/a) (5 a - 7/a) (5 a - 9/a) (5 a - 11/a) (6 a - 7/a)
	\cont\times
	(7 a - 8/a) (7 a - 10/a) (7 a - 12/a) (9 a - 8/a) (11 a - 8/a)
	\cont\Big/ \left(
	9 (35 a - 48/a) (35 a^2 - 97 + 56/a^2) (735 a^2 - 1937 + 1176/a^2)
	\right)
\end{eqnarray}
There were  also the Virasoro descendants of $L_{-2}\vac$ and
of $W_{-4}\vac$.

Having found these fields, we may search for conserved quantities
for the quantum affine theories related to $B_3$. These affine
theories are
$B_3^{(1)}$, $D_5^{(2)}$ and $A_6^{(4)}$.
The duals to these algebras are
$A_5^{(2)}$, $C_3^{(1)}$ and $A_6^{(4)}$ respectively.
\begin{Table}[h]
\begin{center}
\caption{Affine extensions of $B_3$}
\renewcommand{\arraystretch}{1.2}
\vskip .2cm
\begin{tabular}{cccc}
algebra & $B_3^{(1)} $  & $ D_5^{(2)} $ &  $A_6^{(4)}$ \\
root & $(-1,-1,0) $		& $(-1,0,0)$	& $(-2,0,0)$
\end{tabular}
\end{center}
\end{Table}
The conserved quantities for these affine theories are of the form
(\ref{eq.conform}) where $\Phi(z)$ is some field which is polynomial
in the derivatives of the Toda fields.  As above, this charge $\cI$ must
certainly commute with the conformal Hamiltonian, and so we need only
check that the integral $\cI$ obtained from some W-algebra field
commutes with the screening charges corresponding to the extra term in
the potential of the affine theory.

%We systematically searched for such fields $\Phi(z)$ up to
%spin $4$.
%Up to that spin, the fields in the W-algebra are the same as those in
%Table \ref{tab.wfields}.
We systematically searched for such fields $\Phi(z)$ up to spin 6. Up
to that spin the fields in the W-algebra are:
\begin{Table}
\begin{center}
\caption{W-algebra states for $B_3$}
\renewcommand{\arraystretch}{1.2}
\vskip .2cm
\begin{tabular}[c]{|c|c|}
\hline\emptycol{2}\\
  spin 		& state \\
\sline{2}
 $2$		& $L_{-2}\vac $\\
\sline{2}
 $3$		& $L_{-3}\vac $\\
\sline{2}
 $4$		& $L_{-4}\vac\,,\, W_{-4}\vac\,,\, L_{-2}L_{-2} \vac $ \\
\sline{2}
 $5$		& $L_{-5}\vac\,,\, W_{-5}\vac\,,\, L_{-3}L_{-2} \vac $ \\
\sline{2}
		& $L_{-6}\vac\,,\, W_{-6}\vac\,,\, V_{-6}\vac $ \\
 $6$		& $L_{-3}L_{-3}\vac\,,\, L_{-4}L_{-2}\vac $ \\
		& $L_{-2}W_{-4}\vac\,,\, L_{-2}L_{-2}L_{-2} \vac $ \\
\emptycol{2}\\\hline
\end{tabular}
\end{center}
\end{Table}

We find, as expected, that the there is a conserved quantity of spin
one, the momentum
\eq
\int L(z) d\!z \;.
\en
There are no conserved quantities of spin 2, but a conserved quantity
of spin 3, corresponding to a field
$\Phi$ of conformal weight 4, for each of the
affine theories. If we write the field $\Phi$ as
\eq
\Phi_{-4}\vac = W_{-4}\vac + x L_{-2}L_{-2} \vac
\en
then we obtain the conserved quantities for the choices of $x$ below.
\begin{Table}[h]
\begin{center}
\caption{Spin 3 charges}
\vskip .2cm\renewcommand{\arraystretch}{1}
\begin{tabular}{|c|c|c|c|}
\hline \emptycol{4}\\
  & $B_3^{(1)}$	& $D_5^{(2)}$ &	$A_6^{(4)}$ \\
\sline{4}
$x$ &
$ 6 {a^2 (4a^2 - 3) (5a^2 - 9) \over
	(5a^2 - 6) (525a^4 - 1357a^2 + 840)}$ &
$ - 3 {(3a^2 - 5) (3a^2 - 7) (9a^2 - 8) \over
	(3a^2 - 4) (525a^4 - 1357a^2 + 840)}$ &
$ - 12 {(7a^2 - 10) (13a^2 - 7) \over 3675a^4 - 9499a^2 + 5880} $
\\ \emptycol{4}\\
\hline
\end{tabular}
\end{center}
\end{Table}
\noindent
%%%%

For the affine theories $B_3^{(1)}$ and $D_5^{(2)}$ we found further a
conserved quantity of spin 5. If we write the corresponding field
$\Phi$ of spin 6 as
\begin{equation}
\vec\Phi = V_{-6}\vac + x L_{-2}W_{-4}\vac + y L_{-2}L_{-2}L_{-2}\vac +
	z L_{-4}L_{-2}\vac
\end{equation}
then we obtain the conserved quantities for the choices below.
\begin{Table}
\begin{center}
\caption{Spin 5 charges}
\vskip .2cm
\renewcommand{\arraystretch}{1}
\begin{tabular}{|c|c|}
\hline \emptycol{2}\\
  & $B_3^{(1)}$	\\
\sline{2}
$x$ & $
{\frac {10\,a^{2} (5\,a^{2}-3 ) (5\,a^{2}-11 )}
	{ 9\,(5\,a^{2}-6 ) (35\,a^{4}-97\,a^{2}+56
 )}}
$\\ \emptycol{2}\\
$y$ & $
{\frac {(5\,a^{2}-3) (5\,a^{2}-11)
	(1701000\,a^{10}-10548060\,a^{8}+26028772\,a^{6}-
	32003488\,a^{4}+19620096\,a^{2}-4792320 ) }
{3\,(5\,a^{2}-6 )^{2} (35\,a^{2}-48 ) (525\,a^{4}-1357\,a^{2}+840)
	(735\,a^{4}-1937\,a^{2}+1176 ) }}
$\\ \emptycol{2}\\
$z$ & $
	\begin{array}{l}\scriptstyle
	-2\,(5\,a^{2}-3 ) (5\,a^{2}-11 )
	(3307500\,a^{14}-26014275\,a^{12}+87957450\,a^{10}-167212739\,a^{8}
	\\\qquad\scriptstyle+
	194926904\,a^{6}-140484624\,a^{4}+58166784\,a^{2}-10644480 )
	\\ \big/\scriptstyle\left(
	{3\,a^{2} (5\,a^{2}-6 )^{2} (35\,a^{2}-48 )
	(525\,a^{4}-1357\,a^{2}+840) (735\,a^{4}-1937\,a^{2}+1176 )}
	\right)\end{array}
$\\ \emptycol{2} \\
\hline
\hline \emptycol{2}\\
  & $D_5^{(2)}$ \\
\sline{2}
$x$ & $
-{\frac { 5\,(a^{2}-3 ) (3\,a^{2}-7 ) (11\,a^{2}-8 )}
{ 9\,(3\,a^{2}-4 ) (35\,a^{4}-97\,a^{2}+56 )}}
$\\ \emptycol{2}\\
$y$ & $
{\frac {(a^{2}-3) (11\,a^{2}-8)
	(641025\,a^{10}-4699785\,a^{8}+13747319\,a^{6}-19963511\,a^{4}+
	14313192\,a^{2}-4026240 )}
{ 3\,(3\,a^{2}-4 )^{2} (35\,a^{2}-48 ) (525\,a^{4}-1357\,a^{2}+840)
	(735\,a^{4}-1937\,a^{2}+1176 )}}
$\\ \emptycol{2}\\
$z$ & $
	\begin{array}{l}\scriptstyle
	-(a^{2}-3 ) (11\,a^{2}-8 )
		(10087875\,a^{14}-107374050\,a^{12}+487853940\,a^{10}-
		1222163902\,a^{8}
	\\\qquad\scriptstyle+
	1818195337\,a^{6}-1602862752\,a^{4}+774015552\,a^{2}-157731840)
	\\ \big/\scriptstyle\left(
	{6\,a^{2} (3\,a^{2}-4 )^{2} (35\,a^{2}-48 )
	(525\,a^{4}-1357\,a^{2}+840) (735\,a^{4}-1937\,a^{2}+1176 )}
	\right)\end{array}
$\\ \emptycol{2}\\
\hline
\end{tabular}
\end{center}
\end{Table}

The conserved quantities for $A_5^{(2)}, C_3^{(1)}$ and $A_4^{(2)}$
can be obtained by substituting $a \longrightarrow -\sqrt2/a$ in the
above expressions for $B_3^{(1)}, D_5^{(2)}$ and $A_4^{(2)}$.

The spins of the conserved quantities we have found are again
equal to the exponents of the affine Lie algebra, as in the classical
case.
\begin{Table}[h]
\begin{center}
\caption{}\renewcommand{\arraystretch}{1.2}
\vskip .2cm
\begin{tabular}{|c|c|}
\hline\emptycol{2}\\
algebra & exponents \\ \sline{2}
$B_3^{(1)}$ & $i \equiv 1 \bmod 2 $\\
$C_3^{(1)}$ & $i \equiv 1 \bmod 2 $\\
$A_5^{(1)}$ & $i \equiv 1 \bmod 2 $\\
$D_5^{(2)}$ & $i \equiv 1 \bmod 2 $\\
$A_6^{(2)}$ & $i \equiv 1,3,5,9,11,13  \bmod 14 $\\
\emptycol{2}\\\hline
\end{tabular}
\end{center}
\end{Table}

\subsection{$G_2$}

We take the simple roots of $G_2$ to be
\eq
( 0, 1 ), \qquad ( \sqrt{3}/2 , - 3/2 ) \;.
\en
If we consider the conformal theory Toda based on $G_2$ then we can
find fields which commute with the screening charges for the simple
roots of the algebra. Starting with an arbitrary state at each level
we implemented these constraints, and up to conformal weight 6, we
found two independent fields which satisfied them. One was the
Virasoro algebra generator, which corresponded to a state
(\ref{eq.lstat}),
\blank{
\eq
L_{-2}\vac = (H(2,2) (3 A^2 - 6) + 3 H(2,1)^2 A + H(1,2) (9 \sqrt3 A^2 -
10 \sqrt3) + 3 H(1,1)^2 A)/(6 A) \vac
\en
}
and the other was a spin 6 conformal primary field corresponding to
the state
\eq
W_{-6}\vac.
\en
This state is given by an extremely lengthy expression which is an
$O(1/a)$ perturbation of the classical expression.
We normalised the state as
\begin{eqnarray}
\langle 0 | W_6 W_{-6} | 0 \rangle &=&
-576 (a-1/a) (3a-4/a) (a-3/a) (9a-4/a) \cont\times
(4a-5/a) (15a-16/a) (6a-7/a) (7a-8/a) (6a-11/a) \cont\times
(11a-8/a) (7a-10/a) (15a-14/a) (7a-12/a) (9a-7/a) \cont\Big/ \Big(
(168a^2-387+224/a^2) (294a^2-713+392/a^2)
\Big)
\end{eqnarray}

Having found these fields, we may search for conserved quantities
for the quantum affine theories related to $G_2$. These affine
theories are
$G_2^{(1)}$ and $D_4^{(3)}$, corresponding to the extra roots as
below,
\begin{Table}[h]
\begin{center}
\caption{Affine extensions of $G_2$}
\renewcommand{\arraystretch}{1.2}
\vskip .2cm
\begin{tabular}{ccc}
algebra & $G_2^{(1)}$	& $D_4^{(3)}$\\
root & $(-\sqrt3,0)$	& $(- \sqrt 3/2, -1/2)$
\end{tabular}
\end{center}
\end{Table}
\noindent
The conserved quantities $\cI$ for these affine theories
are again of the form (\ref{eq.conform})
We systematically searched for fields $\Phi(z)$ of this form up to
spin $8$.
We again considered arbitrary linear combinations of the W-algebra
fields up to level 8. These are listed below.
\begin{Table}[ht]
\begin{center}
\caption{W-algebra states for $G_2$}
\renewcommand{\arraystretch}{1.2}
\vskip .2cm
\begin{tabular}[c]{|c|c|}
\hline\emptycol{2}\\
  spin 		& state \\
\sline{2}
 $2$		& $L_{-2}\vac $\\
\sline{2}
 $3$		& $L_{-3}\vac $\\
\sline{2}
 $4$		& $L_{-4}\vac\,,\,L_{-2}L_{-2} \vac $ \\
\sline{2}
 $5$		& $L_{-5}\vac\,,\,  L_{-3}L_{-2} \vac $ \\
\sline{2}
 $6$		& $L_{-6}\vac\,,\, W_{-6}\vac \,,\, L_{-3}L_{-3}\vac $\\
		& $L_{-4}L_{-2}\vac \,,\, L_{-2}L_{-2}L_{-2} \vac $ \\
\sline{2}
 $7$		& $L_{-7}\vac\,,\, W_{-7}\vac \,,\, L_{-4}L_{-3}\vac $\\
		& $L_{-5}L_{-2}\vac \,,\, L_{-3}L_{-2}L_{-2} \vac $ \\
\sline{2}
		& $L_{-8}\vac\,,\, W_{-8}\vac \,,\, L_{-4}L_{-4}\vac $\\
 $8$		& $L_{-5}L_{-3}\vac \,,\, L_{-6}L_{-2}\vac \,,\,
			L_{-2}W_{-6}\vac $\\
		& $L_{-3}L_{-3}L_{-2} \vac\,,\, L_{-4}L_{-2}L_{-2}\vac\,,\,
			L_{-2}L_{-2}L_{-2}L_{-2}\vac $ \\
\emptycol{2}\\\hline
\end{tabular}
\end{center}
\end{Table}
\noindent
We found only three such combinations which yielded conserved quantities.
We find, as expected, that the momentum
\eq
\int L(z) d\!z
\en
is a conserved quantity.
We found no conserved quantities at spin less than 5 other than the
momentum. At spin 5 we found conserved quantities of the form
(\ref{eq.conform}), with
\eq
\vec\Phi = (W_{-6} + x L_{-2}L_{-2}L_{-2}  + y L_{-4}L_{-2}) \vac
\en
for both the $G_2^{(1)}$ and $D_4^{(3)}$ theories.
The variables $x,y$ take their values as follows.
\begin{Table}[h]
\begin{center}
\caption{Spin 5 charges}
\vskip .2cm
\begin{tabular}{|c|c|}
\hline \emptycol{2}\\
  & $G_2^{(1)}$ 		\\
\sline{2}
$x$ & $
-{\frac {(6\,a^{2}-11 ) (9\,a^2-4 )
(3048\,a^{6}-9569\,a^{4}+9876\,a^{2}-3328 )}
{ (a^2-1 ) (168\,a^{4}-387\,a^{2}+224 ) (294\,a^{4}-713\,a^{2}+392 ) }}
$\\ \emptycol{2}\\
$y$ & $
{\frac {(6\,a^{2}-11 ) (9\,a^2-4 )
(6888\,a^{8}-22842\,a^{6}+29755\,a^{4}-18864\,a^{2}+4928
 ) (3\,a^{2}-4 )}{2\,a^{2} (a^2-1) (168\,a^{4}-387\,a^{2}+224 )
 (294\,a^{4}-713\,a^{2}+392 ) }}
$\\ \emptycol{2}\\
\hline
\hline \emptycol{2}\\
& $D_4^{(3)}$ \\ \sline{2}
$x$ & $
{\frac { 12\,(a^{2}-3) (11\,a^{2}-8 ) (
1872\,a^{6}-7407\,a^{4}+9569\,a^{2}-4064 )}{
 (3\,a^{2}-4 ) (168\,a^{4}-387\,a^{2}+224 )
(294\,a^{4}-713\,a^{2}+392 ) }}
$\\ \emptycol{2}\\
$y$ & $
-{\frac { 6\,(a^2-1) (11\,a^{2}-8
 ) (8316\,a^{8}-42444\,a^{6}+89265\,a^{4}-91368\,a^{2}+
36736 ) (a^{2}-3 )}{a^{2} (3\,a^{2}-
4 ) (168\,a^{4}-387\,a^{2}+224 ) (294\,a^{4}-713\,a^{2}+392 ) }}
$\\ \emptycol{2}\\
\hline
\end{tabular}
\end{center}
\end{Table}

We found a further conserved quantity at spin 7 of the form
(\ref{eq.conform}), with
\eq
\vec\Phi = L_{-2}W_{-6}\vac + x L_{-2}L_{-2}L_{-2}L_{-2}\vac +
	y L_{-4}L_{-2}L_{-2}\vac + z L_{-6}L_{-2}\vac
\en
for both the $G_2^{(1)}$ and $D_4^{(3)}$ theories.
The variables $x,y$ and $z$ take their values as follows.
\begin{Table}[h]
\begin{center}
\caption{Spin 7 charges}
\vskip .2cm
\begin{tabular}{|c|c|}
\hline \emptycol{2}\\
  & $G_2^{(1)}$ 		\\
\sline{2}
$x$ & $
-{(6\,a^{2}-13)
(30168\,a^{8}-115065\,a^{6}+161672\,a^{4}-99168\,a^{2}+22528) \over
2\,  (a^2-1 ) (168\,a^{4}-387\,a^{2}+224)
 (294\,a^{4}-713\,a^{2}+392 )}
$\\ \emptycol{2}\\
$y$ & $
{3\,(6\,a^{2}-13) (
221256\,a^{12}-1248522\,a^{10}+2895033\,a^{8}-3544800\,a^{6}+2428784\,
a^{4}-887808\,a^{2}+136192) \over
4\,a^2  (a^2-1 ) (168\,a^{4}-387\,a^{2}+224)
 (294\,a^{4}-713\,a^{2}+392 )}
$\\ \emptycol{2}\\
$z$ & $\begin{array}{l}\scriptstyle
-(6\,a^{2}-13) (
7620480\,a^{16}-57057588\,a^{14}+184825566\,a^{12}-339932517\,a^{10}
\\\qquad\scriptstyle+
389929008\,a^{8}-286505024\,a^{6}+131838208\,a^{4}-34736128\,a^{2}+
4014080)
\\ \big/\scriptstyle\left(
5\,a^4 (a^2-1 ) (168\,a^{4}-387\,a^{2}+224 )
 (294\,a^{4}-713\,a^{2}+392 ) \right)
    \end{array}
$\\ \emptycol{2}\\
\hline
\hline \emptycol{2}\\
& $D_4^{(3)}$ \\ \sline{2}
$x$ & $
{6\,(13\,a^{2}-8) (
3168\,a^{8}-18594\,a^{6}+40418\,a^{4}-38355\,a^{2}+13408) \over
 (3\,a^{2}-4 ) (168\,a^{4}-387\,a^{2}+224 )
 (294\,a^{4}-713\,a^{2}+392 )}
$\\ \emptycol{2}\\
$y$ & $
-{3\,(13\,a^{2}-8) (
43092\,a^{12}-374544\,a^{10}+1366191\,a^{8}-2658600\,a^{6}+2895033\,a^
{4}-1664696\,a^{2}+393344) \over
a^2 (3\,a^{2}-4 ) (168\,a^{4}-387\,a^{2}+224 )
 (294\,a^{4}-713\,a^{2}+392 )}
$\\ \emptycol{2}\\
$z$ & $\begin{array}{l}\scriptstyle
12\,(13\,a^{2}-8) (
317520\,a^{16}-3663576\,a^{14}+18539748\,a^{12}-53719692\,a^{10}
\\\qquad\scriptstyle+
97482252\,a^{8}-113310839\,a^{6}+82144696\,a^{4}-33811904\,a^{2}+
6021120)
\\ \big/\scriptstyle\left(
5\,a^4 (3\,a^{2}-4 ) (168\,a^{4}-387\,a^{2}+224 )
 (294\,a^{4}-713\,a^{2}+392 ) \right)
    \end{array}
$\\ \emptycol{2}\\
\hline
\end{tabular}
\end{center}
\end{Table}

\noindent We have checked explicitly that these conserved quantities
obey the required duality properties, that is that they commute with
the terms appearing in the affine Toda Lagrangian for the dual algebra
with $a \to -1/a$.
Again, due to the choice of scale the results for $G_2^{(1)}$ and
$D_4^{(3)}$ transform into each other under
$a\longrightarrow-2/(\sqrt3 a)$.

The spins of the conserved quantities we have found agree with the
spins of the classical conserved quantities, namely that the spins are
equal to the exponents of the affine Lie algebra.

\begin{Table}[h]
\begin{center}
\caption{}\renewcommand{\arraystretch}{1.2}
\vskip .2cm
\begin{tabular}{|c|c|}
\hline\emptycol{2}\\
algebra & exponents \\ \sline{2}
$G_2^{(1)}$ & $i \equiv \pm 1 \bmod 6 $\\
$D_4^{(3)}$ & $i \equiv \pm 1 \bmod 6 $\\
\emptycol{2}\\\hline
\end{tabular}
\end{center}
\end{Table}

\section{Conclusions}

As expected, we found the same number of conserved quantities in these
non-simply laced affine Toda theories as in the classical theories,
and the expressions for these are $O(\hbar)$ deformations of the classical
expressions. We also found the purely quantum property of duality,
that
\eq
\Aff(g,a) = \Aff(g^\vee,-1/a)\;.
\en
We now have the problem of interpretation of this result. There are
two fundamentally different domains of affine Toda
fields theory, namely $a$  purely imaginary and $a$ real. In the first
instance the theory is a massive field theory and it is conjectured
that there is  purely elastic scattering of the particles for $g$
simply-laced. S-matrices have been conjectured for the scatterings
\cite{BCDS1,BCDS4,CMus1}.
In the second regime it is believed that the theory corresponds to a
perturbation of conformal field theory as first considered by
A.~B.~Zamolodchikov \cite{Zamo2}. However here there are problems in
that the Hamiltonian does not obviously describe a unitary time
evolution, and the related problem of null states in the W-algebra
Verma modules which are the particle sectors
(although see \cite{holnew} for a possible route through this
problem).
Here the W-algebras are
well understood for $g$ simply-laced, and it seems very possible that
the Hamiltonian is indeed unitary when acting on the physical subspace
of the free-field Fock space.

An important requirement of the bootstrap approach to the
factorised S-matrix
%approach
was that the
ratios of the particle masses did not change under change of $\b$;
this holds to one loop for the simply-laced theories and for
$A_{2n}^{(2)}$, but not for the
non-simply-laced theories or other twisted theories.

For the non-simply laced affine theories the masses flow as the
coupling constant changes. However, these affine theories are still
thought to be quantum integrable theories since the
particles do not appear to decay to one loop in standard perturbation
theory \cite{Muss1}.
Can we say anything interesting from our calculations here?
We have found that there is a relation between the conserved
quantities of the affine theories for $(\hat g,a)$ and $(\hat
g^\vee,-1/a)$.  One way this should be reflected in the scattering
theory is in the flow of the particle masses.
Since the theories are ``governed'' by their conserved quantities,
then the large $a$ limit of the mass-ratios of the Toda theory based
on $\hat g$ should be the small $a$ limit of the theory based on $\hat
g^\vee$.

The most interesting recent work on the non-simply laced Toda
theories has been the conjecture of the S-matrices for the
$A_{2n-1}^{(2)}, B_n^{(1)}, C_n^{(1)}$ and $D_{n+1}^{(2)}$
theories by Delius et al. \cite{grisetal,grisetalnew}.
They find that the mass ratios do renormalise, and on
closer inspection this is from
those of  the $A_{2n-1}^{(2)}$ theory to those of the $B_{n}^{(1)}$
theory and for $C_n^{(1)}$ to those of $D_{n+1}^{(2)}$.
This indicates that our conjecture on the flow of the masses
is correct,
although their conjectured S-matrices do not respect this duality for
$G_2^{(1)} \longleftrightarrow D_4^{(3)}$.
Further, Monte-Carlo lattice simulations seem
to bear out a flow of masses connecting a theory with its dual, for at
least some of the other affine theories \cite{WW}.

\blank{
There are other interesting possibilities. There has been a recent
flourish of interest in the generalised Drinfeld-Sokolov constructions
of integrable fields theories based on non-maximal embeddings of
$sl(2)\hookrightarrow g$ \cite{PRIN,Int}. These are interacting field
theories of non-abelian Wess-Zumino fields in much the same way that
Toda theory can be thought of as an interacting theory of a $U(1)^l$
WZW fields.  Here the screening charges are of the form
\eq
Q_\b = \int \dbz \Tr( g E_{-\b} g^{-1} \sum_\a E_\a),
\en
where $g$ is a Wess-Zumino field in the group $G_0$ corresponding to
the Lie algebra $g_0$ which is embedded
in $g$ as $g = m_-\oplus g_0\oplus m_+$. Then $\{\a\}$ are the simple
roots of the root lattice of the nilpotent subalgebra $m_+$. The
classical theory is very similar in that there are chiral W-algebras
for the conformal theories of this type, and there are `affine'
theories which may be regarded as perturbations of these. However,
the analysis based on the screening charges seems rather harder in that
the Wess-Zumino field is only now beginning to be understood, and the
intertwiners of the form $Q^i$ need a careful definition of normal
ordering for the Wess-Zumino field.
}

\section{Acknowledgements}

We would like to acknowledge the many
useful conversations we have had
with  P.~Bowcock and J.~Govaerts for a discussion of the first order
Hamiltonian formalism.
GMTW would like to thank E.~Corrigan, P.~Mansfield and R.~A.~Weston
for many helpful discussions, and
P.~Dorey and H.~Braden for explaining some aspects of their own work.
GMTW is grateful to the SERC for a research
assistantship.
HGK would like to thank the Studienstiftung des deutschen Volkes
for a research studentship.
This work was initiated while the authors were visiting the Institute
for Theoretical Physics at the University of California at Santa Barbara,
where they were supported in part by the National
Science Foundation under Grant No.~PHY89-04035.
\par\pagebreak
\newpage

%\bibliography{bib2,bib1}

\begin{thebibliography}{10}

\bibitem{Mans2}
P.~Mansfield,
\newblock {\em Conformally Extended Toda Theories},
\newblock Phys. Lett. B242 (1990) 387.

\bibitem{FLuk1}
V.~A. Fateev and S.~L. Luk'yanov,
\newblock {\em The models of Two-Dimensional Conformal Quantum Field Theory
  with {$Z_n$} Symmetry},
\newblock Int. J. Mod. Phys. A3 (1988) 507.

\bibitem{FLuk2}
V.~A. Fateev and S.~L. Luk'yanov,
\newblock {\em Additional Symmetries and Exactly-Soluble Models in
  Two-Dimensional Conformal Field Theory},
\newblock Sov. Sci. Rev. A15 (1990) 1.

\bibitem{KWat3}
H.~G. Kausch and G.~M.~T. Watts,
\newblock {\em Quantum Toda theory and the Casimir algebra of $B_2$ and $C_2$},
\newblock Durham University Preprint DTP-91-35 (1991),
\newblock to appear in Int. J. Mod. Phys.~A.

\bibitem{Destetal}
C.~Destri, H.~J. de~Vega and V.~A. Fateev,
\newblock {\em The Exact S-Matrices associated to non-simply laced affine Toda
  Field Theories: The $B_n^{(1)}$ and $C_n^{(1)}$ cases},
\newblock Phys. Lett. b256 (1990) 173.

\bibitem{fren1}
E.~V. Frenkel,
\newblock {\em Affine Kac-Moody Algebras at the Critical Level and Quantum
  Drinfeld-Sokolov Reduction},
\newblock PhD thesis, Harvard University, 1991.

\bibitem{DSok2}
V.~G. Drinfel'd and V.~V. Sokolov,
\newblock {\em Lie Algebras and Equations of Korteweg-de Vries Type},
\newblock J. Sov. Math. 30 (1985) 1975.

\bibitem{EYan1}
T.~Eguchi and S.-K. Yang,
\newblock {\em Deformations of Conformal Field Theories and Soliton Equations},
\newblock Research Institute for Fundamental Physics, Kyoto University RIFP-797
  (1987).

\bibitem{Pall1}
L.~Palla,
\newblock {\em Perturbed $W$ algebras and affine Toda theories},
\newblock Nucl. Phys. B341 (1990) 714.

\bibitem{FFre4}
B.~L. Feigin and E.~V. Frenkel,
\newblock {\em Free Field resolutions and affine Toda theory},
\newblock Harvard University Preprint  (1991).

\bibitem{BCDS1}
H.~W. Braden, E.~Corrigan, P.~E. Dorey and R.~Sasaki,
\newblock {\em Extended Toda Field Theory and Exact S-Matrices},
\newblock Phys. Lett. B227 (1989) 411.

\bibitem{BCDS2}
H.~W. Braden, E.~Corrigan, P.~E. Dorey and R.~Sasaki,
\newblock {\em Aspects of Perturbed Conformal Field Theory, Affine Toda Field
  Theory and Exact S-Matrices},
\newblock University of Durham Preprint UDCPT-89-35 (1989).

\bibitem{grisetal}
G.~W. Delius, M.~T. Grisaru and D.~Zanon,
\newblock {\em Exact S-Matrices for the non-simply-laced affine Toda theories
  $a^{(2)}_{2n-1}$},
\newblock Cern preprint CERN-TH.6333/91 (1991).

\bibitem{Cole1}
S.~Coleman,
\newblock {\em Quantum Sine Gordon Theory as the massive Thirring Model},
\newblock Phys. Rev. D11 (1975) 2088.

\bibitem{Mansfield.earlier.toda}
P.~Mansfield,
\newblock {\it Solution of Toda systems}, \newblock Nucl. Phys. B208 (1982)
  277; {\it Light-cone quantisation of the Liouville and Toda field theories},
  \newblock B222 (1983) 419.

\bibitem{Godd1}
P.~Goddard,
\newblock {\em Meromorphic Conformal Field Theory},
\newblock {\em in\/:} Infinite Dimensional Lie Algebras and Lie Groups,  ed.
  V.~G. Kac, World Scientific, 1989,
\newblock CIRM-Luminy July conference on Infinite dimensional Lie Algebras and
  Lie Groups, Marseille 1988.

\bibitem{GOli1}
P.~Goddard and D.~Olive,
\newblock {\em Kac-Moody and Virasoro Algebras in Relation to Physics},
\newblock Int. J. Mod. Phys. A1 (1986) 303.

\bibitem{deveganew}
C.~Destri and H.~J. de~Vega,
\newblock {\em New exact results in affine Toda field theories: Free energy and
  wave function renormalizations},
\newblock Nucl. Phys. B358 (1991) 251.

\bibitem{Kac1}
V.~Kac,
\newblock {\em Infinite dimensional Lie algebras},
\newblock Cambridge University Press, 1985.

\bibitem{KWat1}
H.~G. Kausch and G.~M.~T. Watts,
\newblock {\em A Study of W-algebras using Jacobi identities},
\newblock Nucl. Phys.  (1991) 740.

\bibitem{BCDS4}
H.~W. Braden, E.~Corrigan, P.~E. Dorey and R.~Sasaki,
\newblock {\em Affine Toda Field Theory and Exact S Matrices},
\newblock Nucl. Phys. B338 (1990) 689 .

\bibitem{CMus1}
P.~Christe and G.~Mussardo,
\newblock {\em Integrable Systems away from criticality: The Toda Field Theory
  and S Matrix of the Tricritical Ising model},
\newblock Nucl. Phys. B330 (1990) 465.

\bibitem{Zamo2}
A.~B. Zamolodchikov,
\newblock {\em Integrable Field Theory from Conformal Field Theory},
\newblock {\em in\/:} Advances in Pure Mathematics 19, Integrable Systems in
  Quantum Field Theory and Statistical Mechanics,  eds. M.~Jimbo, T.~Miwa and
  A.~Tsuchiya, pages 641--674, Kinokuniya Company, Tokyo, 1989.

\bibitem{holnew}
T.~J. Hollowood,
\newblock {\em Quantum solitons in affine Toda field theories},
\newblock Princeton preprint PUPT-1286 (1991).

\bibitem{Muss1}
G. Mussardo, private communication.

\bibitem{grisetalnew}
G.~W. Delius, M.~T. Grisaru and D.~Zanon,
\newblock {\em Exact S-Matrices for non-simply-laced affine Toda theories},
\newblock Cern preprint CERN-TH.6337/91 (1991).

\bibitem{WW}
R.~A.~Weston and G.~M.~T.~Watts, work in progress.

\end{thebibliography}

\end{document}